\documentclass{Interspeech}
\interspeechcameraready 

\title{The Text-to-speech in the Wild (TITW) Database}

\author[affiliation={1}]{Jee-weon}{Jung$^\dagger$}
\author[affiliation={2}]{Wangyou}{Zhang}
\author[affiliation={1,3}]{Soumi}{Maiti}
\author[affiliation={4}]{Yihan}{Wu}
\author[affiliation={5}]{Xin}{Wang}
\author[affiliation={6}]{\\Ji-Hoon}{Kim}
\author[affiliation={7}]{Yuta}{Matsunaga}
\author[affiliation={8}]{Seyun}{Um}
\author[affiliation={1}]{Jinchuan}{Tian}
\author[affiliation={1}]{Hye-jin}{Shim}
\author[affiliation={9}]{\\Nicholas}{Evans}
\author[affiliation={6}]{Joon Son}{Chung}
\author[affiliation={10}]{Shinnosuke}{Takamichi}
\author[affiliation={1}]{Shinji}{Watanabe}

\affiliation{}{Carnegie Mellon University}{USA}
\affiliation{}{Shanghai Jiao Tong University}{China}
\affiliation{}{Meta}{USA\\}
\affiliation{}{Renmin University of China}{China}
\affiliation{}{National Institute of Informatics}{Japan\\}
\affiliation{}{Korea Advanced Institute of Science and Technology}{South Korea}
\affiliation{}{University of Tokyo}{Japan}
\affiliation{}{Yonsei University}{South Korea}
\affiliation{}{EURECOM}{France}
\affiliation{}{Keio University}{Japan}

\email{jeeweonj@ieee.org}

\usepackage{amsmath,graphicx}
\usepackage{cleveref}
\usepackage{xpatch,xcolor}
\usepackage{makecell,booktabs}
\usepackage{amsmath,amssymb}
\usepackage{mathtools}
\usepackage{multicol,multirow}
\usepackage{array}
\usepackage{tabularx}
\usepackage{caption}
\usepackage{subcaption}
\usepackage{enumitem,kantlipsum}
\usepackage{cite}
\usepackage{url}
\usepackage{hyperref}

\hypersetup{
    colorlinks=true,
    linkcolor=purple,
    filecolor=magenta,      
    urlcolor=purple,
}
\newcolumntype{Y}{>{\centering\arraybackslash}X}
\newcolumntype{C}{ >{\centering\arraybackslash} m{4cm} }

\newcommand{\newpara}[1]{\vspace{2pt}\noindent\textbf{#1}}

\usepackage{comment}
\usepackage{enumitem,amssymb}
\newlist{todolist}{itemize}{2}
\setlist[todolist]{label=$\square$}

\makeatletter
\renewcommand{\section}{\@startsection{section}{1}{0pt}%
  {6pt} 
  {3pt} 
  {\normalfont\normalsize\bfseries}}  

\renewcommand{\subsection}{\@startsection{subsection}{2}{0pt}%
  {2pt} 
  {1pt} 
  {\normalfont\normalsize\bfseries}}  

\renewcommand{\subsubsection}{\@startsection{subsubsection}{3}{0pt}%
  {2pt} 
  {1pt} 
  {\normalfont\normalsize\itshape}}  
\makeatother

\begin{document}
\maketitle

\begin{abstract}
Traditional Text-to-Speech (TTS) systems rely on studio-quality speech recorded in controlled settings. Recently, an effort known as ``noisy-TTS training" has emerged, aiming to utilize in-the-wild data. However, the lack of dedicated datasets has been a significant limitation. We introduce the TTS In the Wild (TITW) dataset, which is publicly available\footnote{\url{https://huggingface.co/datasets/jungjee/titw}}, created through a fully automated pipeline applied to the VoxCeleb1 dataset. It comprises two training sets: TITW-Hard, derived from the transcription, segmentation, and selection of raw VoxCeleb1 data, and TITW-Easy, which incorporates additional enhancement and data selection based on DNSMOS. State-of-the-art TTS models achieve over 3.0 UTMOS score with TITW-Easy, while TITW-Hard remains difficult showing UTMOS below 2.8. Beyond TTS, TITW’s unique design, leveraging an automatic speaker recognition dataset, strengthens ethical efforts to counteract malicious use of TTS models by supporting tasks such as speech deepfake detection.

\end{abstract}
\renewcommand{\thefootnote}{\fnsymbol{footnote}}
\footnotetext[2]{Currently at Apple.}
\renewcommand{\thefootnote}{\arabic{footnote}}
\section{Introduction}
\label{sec:intro}
Generative speech technology is evolving rapidly, driven in part by advances in diffusion models, speech codecs, and speech-language modeling methodologies~\cite{le2024voicebox,kim2024p,nguyen2024fregrad,zhang2023speechtokenizer,yang2023uniaudio}. Among these advancements, Text-to-Speech (TTS) systems have made remarkable progress, with recent models capable of generating speech that is nearly indistinguishable from human speech in terms of intelligibility and naturalness. Notably, while traditional TTS systems required minutes of studio-recorded target speaker data, modern systems can now operate effectively with just a few seconds of such data~\cite{lee2022pvae,kharitonov2023speak,kimclam}.

In terms of TTS training, however, studio-quality data remains the \textit{de facto} standard, despite its limitations. 
Studio recordings, while superior in audio quality, lack diversity, variability, and scalability. 
In-the-wild speech data, on the other hand, offers exposure to real-world variability, greater speaker diversity, and nearly unlimited scalability.
These benefits are particularly valuable for underrepresented languages, as studio-based data is often limited, hindering the democratization of speech technology. Being able to utilize publicly available sources like YouTube for TTS training could revolutionize the field, enabling the curation of target language data from accessible, diverse, and abundant resources.  

\begin{figure}
    \centering
    \vspace{10pt}
    \includegraphics[width=\columnwidth]{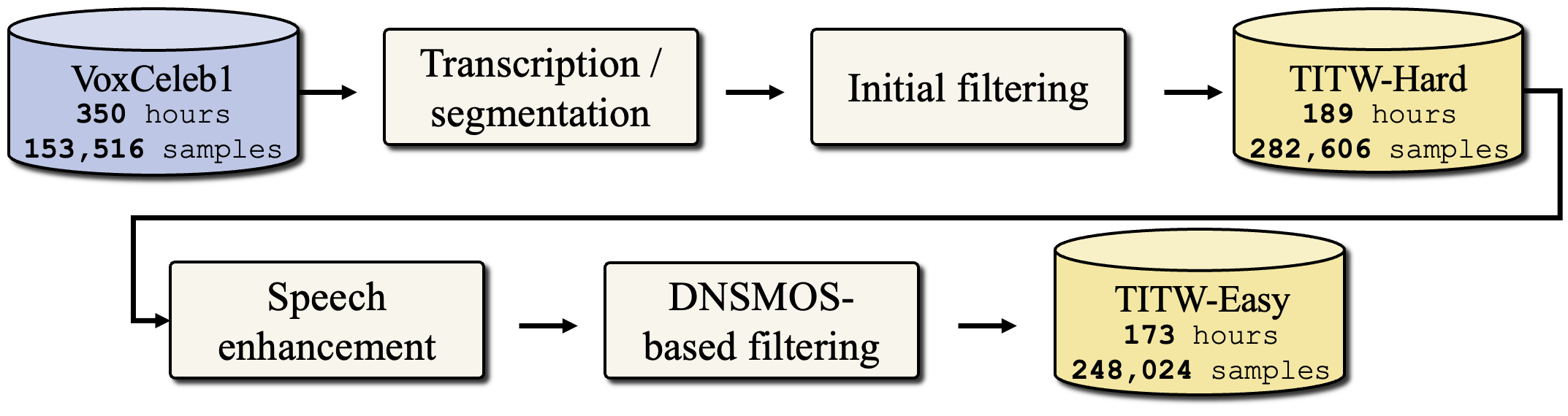}
    \caption{
    Fully automated Text-To-Speech Synthesis In The Wild (TITW) processing pipeline. 
    The pipeline incorporates transcription, segmentation, data selection, enhancement, and filtering based on DNSMOS.
    The TITW dataset comprises two editions: \textbf{TITW-easy}, which can be used to successfully train latest TTS systems, and \textbf{TITW-hard}, where the quality is too low to train TTS systems with current technology but aims to serve the training of more advanced future TTS systems.}

    \label{fig:pipeline}
\end{figure}

Efforts to train TTS systems with lower-quality, in-the-wild data -- often termed ``noisy-TTS training" -- have shown promise. While earlier works have reported that TTS models cannot be trained with low-quality data~\cite{yamagishi2010thousands,karhilaNoise2014}, more recent studies such as DenoiSpeech~\cite{zhang2021denoispeech} and MQTTS~\cite{chen2023vector} demonstrate that competitive performance can be achieved using non-studio-quality data, such as podcasts and YouTube audio. 
The ASVspoof 5 challenge~\cite{wang2024asvspoof} further supports this shift, demonstrating the potential of real-world sources like audiobooks for high-quality synthetic speech. However, the field of noisy-TTS training lacks standardized datasets, protocols, and benchmarks, with most studies relying on private data or artificially noised studio recordings. This gap hinders progress and reproducibility in the field.

To this end, we introduce the Text-To-Speech Synthesis In The Wild (TITW) database, which is publicly available. 
It is constructed using the VoxCeleb1 database~\cite{nagrani2017voxceleb}, a large collection of YouTube speech data, and processed through an automated pipeline involving transcription, segmentation, and enhancement. 
This pipeline eliminates the need for manual processing, making the dataset scalable and accessible. TITW includes two training sets: TITW-Hard, comprising 189 hours of speech derived from raw VoxCeleb1 data with minimal processing, and TITW-Easy, a refined subset of 173 hours enhanced using DNSMOS-based selection and additional processing. We also propose standardized evaluation protocols and benchmarks to facilitate reproducible research. Our experiments demonstrate that four contemporary TTS models can be successfully trained using TITW, showcasing its practical utility.

The selection of VoxCeleb1 as a source dataset, originally developed for automatic speaker recognition, offers unique and novel benefits to TITW.
Recent TTS systems, capable of producing speech nearly indistinguishable from human voices, have raised concerns about malicious use and its potential to damage society~\cite{mai2023warning}.
By training TTS models on TITW, the resulting synthetic speech can be paired with human speech in TITW to support research in speech deepfake detection and spoofing-robust automatic speaker verification~\cite{jung2022sasv}. This application of TITW not only advances TTS research but also contributes to developing robust countermeasures against the misuse of synthetic voices.
This dual benefit stems from the dataset’s guarantee that all speech samples feature only single-speaker audio. Consequently, our work underscores a broader impact, fostering both innovation and ethical considerations in the field of generative speech technology.

\section{Related works}
\label{sec:related}
Numerous databases have been used for training TTS models. 
Legacy databases such as CMU ARCTIC~\cite{kominek2004cmuarctic} and VCTK~\cite{yamagishiCSTR2019} were carefully designed and curated.  They contain phonetically-balanced utterances, all recorded in highly controlled acoustic environments. 
Due to the high cost of recording, these and similar databases typically include data from a single speaker or a small number of speakers. 
The speech data they contain is generally neutral in terms of emotions and expressiveness. These databases were widely used for training speaker-dependent and multiple-speaker legacy TTS systems (e.g., unit-selection~\cite{hunt1996unit} and HMM-based~\cite{tokuda2013speech}).

The revolution in deep-learning-based TTS systems called for larger-scale datasets. 
Datasets like LJSpeech~\cite{ljspeech17}, Multi-lingual Librispeech~\cite{pratap2020mls} and LibriTTS~\cite{zen2019libritts}, which are sourced from LibriVox audiobooks, are not recorded in studio-quality environments.
LJSpeech contains twenty-four hours of audiobook recordings but from a single speaker while the other two feature a greater number of speakers. 
They have been widely used to train deep-learning-based TTS systems~\cite{wang2023neural, shennaturalspeech}. 
Their adoption marks a shift towards using training data collected in less controlled conditions. 
Even so, this data still falls short of capturing the diversity in speaker style and acoustic conditions found in truly ``in-the-wild'' scenarios; the signal-to-noise ratio remains high and utterances are generally well-enunciated.\footnote{See Librivox documentation and guidelines for recording an audiobook  \url{https://librivox.org/pages/about-recording}.}

The TITW database introduce in this paper aims to support research in overcoming data constraints, often referred to as noisy-TTS training. 
We envisage TTS systems that can be trained successfully using speech data collected in uncontrolled conditions. 
We see two avenues for such research.
The first, most challenging direction involves the use of training data collected in the wild without manual human intervention, relying solely on automatic transcriptions, segmentation, and data selection based on heuristics. 
The second direction involves the use of a subset of data after applying additional speech enhancement and data selection based on speech quality.  
TITW contains in-the-wild recordings of interviews, podcasts, and more, all posted to social media, making it, to our best knowledge, one of the first of its kind.\footnote{We recognize EMILIA~\cite{he2024emilia} as the most similar, parallel work. However, the goals of the two works differ. EMEILIA focuses on developing a data processing pipeline that yields high-quality data from in-the-wild data. Therefore, it strives to provide the highest achievable quality. TITW is designed to foster research in the training of TTS systems using more noisy and real-world data. Hence we provide not only TITW-Easy which can be used for the training of contemporary TTS systems, but also TITW-Hard to challenge the development of future systems.}

\begin{figure}
    \centering
    \includegraphics[width=\columnwidth,trim={0 0 3cm 0},clip]{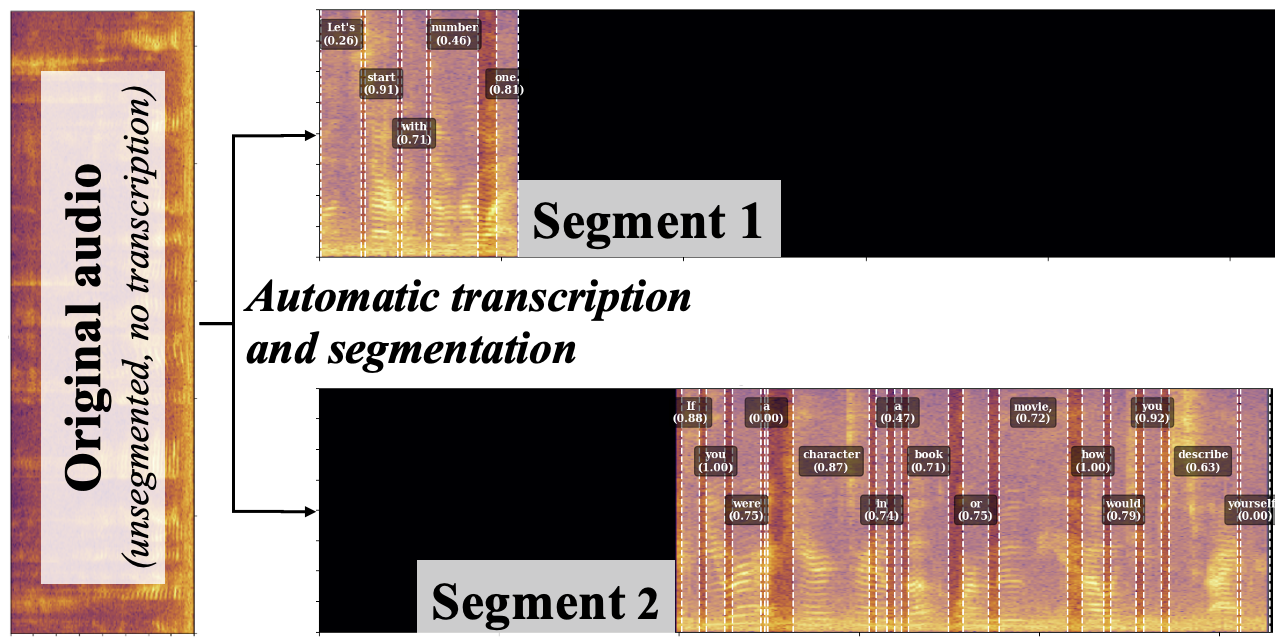}
    \caption{
    An example of the transcription and segmentation in TITW automatic pipeline. 
    A randomly selected utterance from VoxCeleb1 goes through our transcription and segmentation pipeline, deriving two segments.
    A segment in the middle is deleted because it is a non-speech segment over 500ms.
    }
    \label{fig:trans_seg}
    \vspace{-20pt}
\end{figure}

\section{TITW}
\label{sec:dataset}
For several reasons, we selected VoxCeleb1~\cite{nagrani2017voxceleb}, which contains speech from $1,251$ speakers, as the source data for TITW. 
Firstly, VoxCeleb1 is itself sourced from the wild, specifically, YouTube, spanning diverse acoustic environments.
Secondly, as a dataset for automatic speaker recognition, each utterance is from a single speaker.
Lastly, by selecting VoxCeleb1 as a source dataset, TTS systems trained on TITW can contribute to future research in speech deepfake detection and spoofing-robust automatic speaker verification, especially requiring TTS researchers' attention to safeguard the rapidly advancing speech generation technology from malicious usage.
SpoofCeleb~\cite{jung2024spoofceleb} exemplifies this effort, where synthesized (spoofed) utterances generated by 23 TTS systems trained on TITW-Easy are used to create a dataset for speech deepfake detection and spoofing-robust automatic speaker verification.

\subsection{Transcription and segmentation}
\label{ssec:voxceleb_preprocessing}

We first transcribe and segment the utterances using pretrained models and empirically derived heuristics, ensuring the process is fully automatic without human intervention.
Figure~\ref{fig:trans_seg} displays an example of a VoxCeleb1 utterance, transcribed at the word-level and then segmented into two.

\newpara{Transcription.} Since TTS training typically requires paired speech and text data, we generated transcriptions for the entire VoxCeleb1 corpus. 
For the sake of scalability and reproducibility in future projects in various languages, we generated transcriptions automatically using pretrained automatic speech recognition (ASR) models.
We used the WhisperX~\cite{WhisperX-Bain2023} toolkit to generate transcriptions with word-level timestamps. WhisperX incorporates the OpenAI Whisper Large v2 model~\cite{Whisper-Radford2023} for transcription and another phoneme-based ASR model for word-level alignment.
We additionally employed the OSWMv3 speech foundation model~\cite{OWSMv3-Peng2024} and transcribed the data in parallel. 
The transcriptions from the OWSMv3 model served to verify transcription accuracy.

\newpara{Segmentation.} We divide each sample into isolated segments using Voice Activity Detection (VAD) embedded in WhisperX~\cite{WhisperX-Bain2023}.
Practically, whenever a non-speech periods exceed $500$ms, we trim the silence and split it into two segments.
This segmentation rule was developed through empirical, iterative investigations.
Initial attempts to train TTS models with unsegmented data failed, revealing that excessively long silences within training samples were a significant issue.
This procedure results in approximately 280k transcribed speech segments.

\subsection{Data selection}
\label{ssec:filtering_se}
In order to maximize the ``wildness'' of the dataset, our initial investigations did not apply any data selection mechanism when composing the training set for TTS. 
However, we empirically found that attempts to train TTS models were unsuccessful due to the data being excessively noisy, including issues like mistranscription. 
Consequently, we developed four heuristically defined rules for data selection.
These heuristics emerged from iterative efforts to train TTS models with filtered data. 
If any of the following conditions are met, the data is removed and discarded from further consideration:
\begin{itemize}
    \item The language is not from target language,  in this case, English. To simplify TTS training, we use Whisper's language recognition capability to detect and remove utterances in languages other than English.
    Multilingual extensions are left for future work.
    \item The segment duration is shorter than 1 second or longer than 8 seconds. 
    Empirical evidence suggests that using utterances with such a semi-consistent duration benefits training stability.
    \item The per-word duration is longer than 500 ms. 
    The typical speaking rate is in the order of 2 words per second. 
    Outliers often correspond to emotional or pathological speech, or long intervals of non-speech, all of which can destabilize TTS training and are hence removed. 
    \item The automatic transcription is empty. Such cases indicate a non-speech segment or ASR failure.  In either case, they cannot be used for TTS training and are removed.\footnote{Note that despite the application of these data selection heuristics, TITW retains a higher level of variability and naturalness compared to most existing corpora.}
\end{itemize}

The application of transcription, segmentation and data selection results in the ``\textit{TITW-Hard}'' database.  Since the raw data is collected from videos posted to social media, utterances in the TITW-Hard database still contain background noise or low-quality speech.
Preliminary experiments have revealed that the training of TTS models with TITW-Hard data is extremely challenging; most attempts failed to converge even after applying the aforementioned data selection heuristics.

\begin{figure}
    \centering
    \includegraphics[width=\columnwidth]{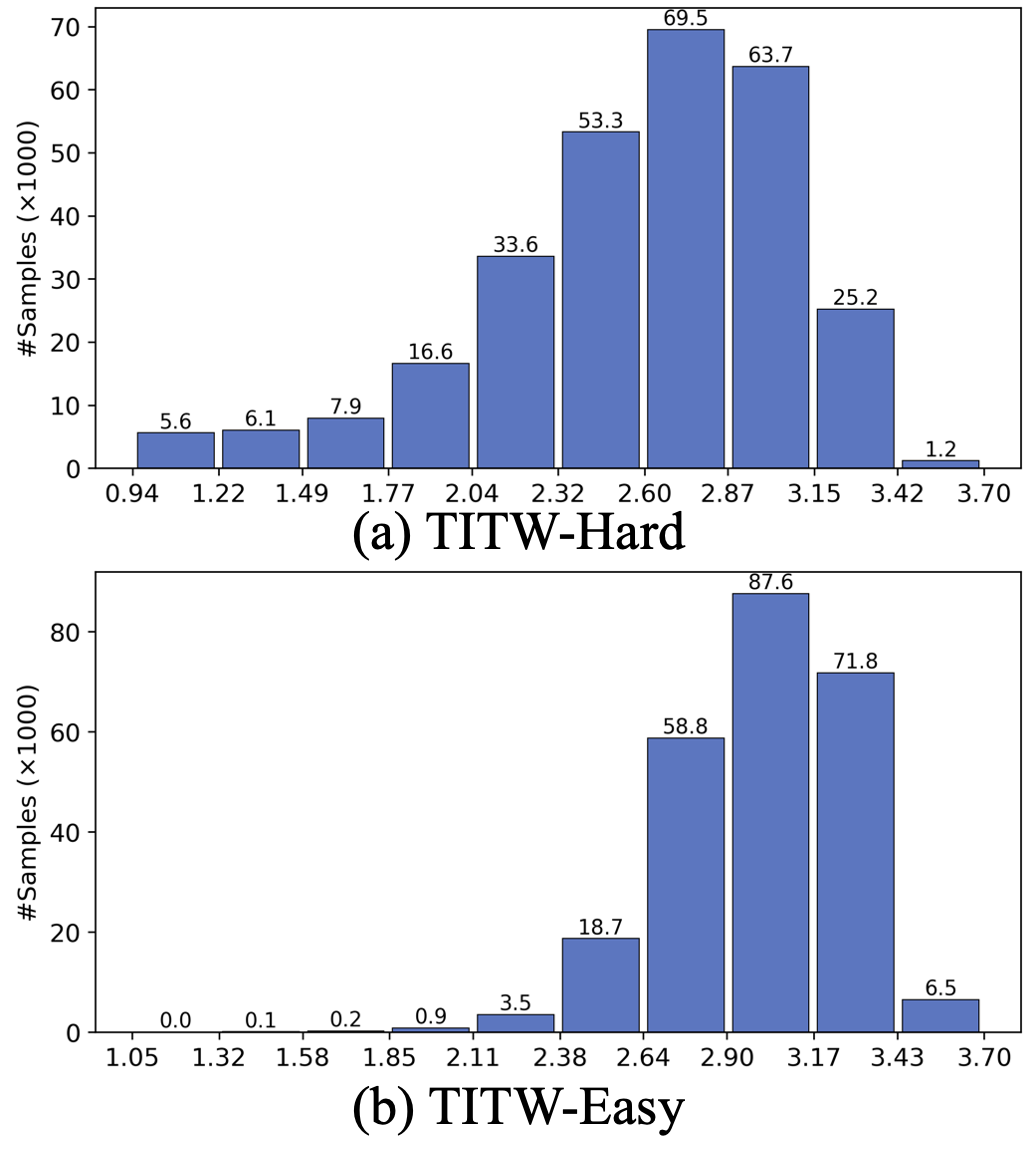}
    \caption{
    Histograms of samples in the TITW-Easy and -Hard sets using DNSMOS overall score shown in the x-axis.
    Even with data selection heuristics discussed in Section~\ref{ssec:filtering_se}, training with TITW-Hard remains extremely challenging.
    }
    \label{fig:data_quality}
    \vspace{-15pt}
\end{figure}
\begin{table}[t!]
    \caption{
    Text-To-Speech Synthesis In The Wild (TITW) statistics. Both sets involve $1,251$ speakers.
    }
    \label{tab:db_stat}
    \centering
    \resizebox{\linewidth}{!}{
    \begin{tabular}{lcccccc}
    \toprule
    & \# samples & Avg dur (s) & Tot dur (h) & Avg \# words\\
    \toprule
    TITW-Easy & $282,606$ & $2.42$ & $189$ & $10.84$\\
    TITW-Hard & $248,024$ & $2.51$ & $173$ & $10.55$\\
    \bottomrule
    \end{tabular}%
    }
    \vspace{-10pt}
\end{table}
\begin{table}[t]
    \caption{
    Speech quality of the TITW-Easy and -Hard sets. WER is calculated using OWSMv3~\cite{OWSMv3-Peng2024}. TITW remains significantly more challenging than VCTK or even EMILIA with a DNSMOS of 3.20 and 3.26.
    }
    \label{tab:sqa_bona}
    \centering
    \begin{tabular}{lcccc}
        \toprule
         &  UTMOS & DNSMOS & WER (\%)\\
        \toprule
        TITW-Hard & 3.00 & 2.38 & 9.30\\
        TITW-Easy & 3.32 & 2.78 & 9.10\\
        \bottomrule
    \end{tabular}
    \vspace{-15pt}
\end{table}

\subsection{Enhancement and DNSMOS-based further data selection}
\label{ssec:additional_filtering}
Given the challenges of training TTS models using the TITW-Hard database, we created a second, relatively less challenging dataset named ``\textit{TITW-Easy}.''\footnote{We believe that future TTS models and training schematics will enable successful training with TITW-Hard. Nonetheless, we introduce TITW-Easy, which contemporary state-of-the-art TTS architectures and training schemes can effectively utilize, as a stepping stone towards research in noisy-TTS training.} 
First, we apply a pretrained speech enhancement model, DEMUCS~\cite{Denoiser-Dfossez2020}\footnote{\url{https://github.com/facebookresearch/denoiser}} to reduce additive, background noise.\footnote{The application of denoising does not compromise our objective to train TTS models with automatically collected data from the wild -- the entire pipeline remains automated, reproducible, and scalable.}
We then apply a second round of data selection, this time to the enhanced data.  
This is done by estimating DNSMOS scores~\cite{DNSMOS-Reddy2021,DNSMOS_P835-Reddy2022} for each utterance.
Then, all utterances for which the DNSMOS score is below a threshold of 3.0 are removed. An exception is made for segments from speakers included in the generation protocol (Section~\ref{sec:eval}).

Figure~\ref{fig:data_quality} presents histograms of DNSMOS scores for both the TITW-Hard and TITW-Easy databases. Is is clearly shown that most of the low-quality samples with low DNSMOS scores have been filtered out.
Table~\ref{tab:db_stat} presents statistics and Table~\ref{tab:sqa_bona} further details the UTMOS, DNSMOS overall, and Word Error Rate (WER) of TITW-Easy and -Hard providing a comprehensive measure of the overall quality and intelligibility. WER was calculated by comparing TITW's transcript with OWSMv3.
The low DNSMOS scores confirm that both TITW training sets retain their challenging nature.

\begin{table*}[t]
    \caption{
    Speech quality of the segments generated from the baselines on the TITW-KSKT and -KSUT protocols.
    All models were trained using the TITW-Easy data. MCD is only applicable for TITW-KSKT where it has the reference samples.
    }
    \vspace{-5pt}
    \label{tab:easyset}
    \centering
    \resizebox{\textwidth}{!}{
    \begin{tabular}{l|cccc|cccc}
        \toprule
        \multirow{2}{*}{System} & \multicolumn{4}{c}{TITW-KSKT} & \multicolumn{4}{c}{TITW-KSUT}\\
        & MCD$\downarrow$ & UTMOS$\uparrow$ & DNSMOS$\uparrow$ & WER (\%)\,$\downarrow$ & MCD$\downarrow$ & UTMOS$\uparrow$ & DNSMOS$\uparrow$ & WER (\%)\,$\downarrow$\\
        \toprule
        TransformerTTS-ParallelWaveGAN & 11.68 & 2.06 & 2.50 &  24.90 & N/A & 1.79 & 2.54 & 107.90 \\
        GradTTS-DiffWave & 6.76 & 2.18 & 2.39 & 11.90 & N/A & 2.30 & 2.54 & 54.00 \\
        VITS & 8.61& 2.77 & 2.74 & 53.00 & N/A & 2.78 & 2.81 & 120.50 \\
        MQTTS & 6.99 & 3.08 & 2.83 & 23.30 & N/A & 3.20 & 2.94 & 67.10\\
        \bottomrule
    \end{tabular}}
    \vspace{-15pt}
\end{table*}
\begin{table}[t]
    \caption{
    Comparative results of the identical TTS system being trained with TITW-Easy and TITW-Hard data.
    Metrics are reported using the TITW-KSKT protocol.
    ``GTmel'' uses ground truth mel-spectrograms in place of GradTTS to solely measure vocoder's performance. 
    }
    \label{tab:hardset}
    \centering
    \resizebox{\linewidth}{!}{
    \begin{tabular}{lc|cccc}
        \toprule
        System & Train & MCD & UTMOS & DNSMOS & WER\\
        \toprule
        GTmel-DiffWave & Easy & 5.05 & 2.63 & 2.68 & 11.90\\
        GTmel-DiffWave & Hard & 4.97 & 2.24 & 2.30 & 12.20 \\
        \midrule
        GradTTS-DiffWave & Easy & 6.76 & 2.18 & 2.39 & 11.90\\
        GradTTS-DiffWave & Hard & 8.23 & 1.29 & 1.47 & 26.20 \\
        \toprule
        VITS & Easy & 8.61 & 2.77 & 2.74 & 53.00\\
        VITS & Hard & 9.06 & 2.48 &	2.69 & 59.50\\
        \bottomrule
    \end{tabular}}
    \vspace{-20pt}
\end{table}

\section{Evaluation and benchmarking}
\label{sec:eval}
Once a TTS model is trained using the TITW database, it can be evaluated with one of the two protocols for generating new synthetic speech.

\newpara{TITW-KSKT (Known Speaker, Known Text)} is designed to generate synthetic speech for speakers and text that are \emph{both} present in the TITW-Hard and TITW-Easy training sets.   
Both sets of speakers and text are randomly extracted from those used in the VoxCeleb1-O automatic speaker verification evaluation protocol.
Consequently, the number of speakers here matches that of the VoxCeleb1-O protocol, at $40$. 
However, due to our data preparation processes outlined in Section~\ref{sec:dataset}, the number of segments has increased from $4,708$ to $9,113$.

\newpara{TITW-KSUT (Known Speaker, Unknown Text)} aims to generate synthetic speech with text that is unseen in both the TITW-Hard and TITW-Easy datasets. We employ two text sources for this:
The first is the Rainbow Passage~\cite{fairbanks1960voice}, which covers many English sounds and their combinations. It has been used widely in other data collection efforts, for example the VCTK corpus~\cite{yamagishiCSTR2019}. 
The second is a set of Semantically Unpredictable Sentences (SUS)~\cite{benoit1996sus} selected from past Blizzard challenges~\cite{king2014measuring}. 
In total, there are $200$ different text samples ($31$ from The Rainbow Passage and $169$ from the set of SUS).
With the same set of 40 speakers, the protocol requires the generation of $8,000$ ($=40\times200$) synthetic utterances.

\section{Experiments}
\label{sec:exps}

\subsection{Metrics} 
We adopt four metrics to assess the quality of generated synthetic speech: (1) Mel Cepstral Distortion~\cite{kubichek1993mel} (MCD) measures the spectral similarity between synthesized and natural speech; (2) UTMOS~\cite{saeki2022utmos} estimates the overall speech quality; (3) DNSMOS~\cite{DNSMOS_P835-Reddy2022} also estimates the overall quality, including aspects such as noise reduction; (4) the ASR WER, measured using the OpenAI Whisper-Large model~\cite{radford2023robust}, quantifies the intelligibility of speech by measuring transcription errors.
We use all four metrics as different proxies for speech quality.
In practice, we use the VERSA toolkit to compute all four metrics~\cite{shi2024codec}.

\subsection{TTS training data}
To provide reference, we first compared the two TITW datasets with others commonly used for TTS training.
Results presented in Table~\ref{tab:sqa_bona} indicate that the TITW-Easy dataset surpasses the TITW-Hard dataset in terms of quality as intended.
As expected, speech samples in both TITW datasets remain more challenging than those used typically for TTS training. 
DNSMOS scores of TITW-Easy and -Hard are 2.78 and 2.38 while those of  VCTK~\cite{veaux201x_cstr_vctk}, MLS~\cite{pratap2020mls}, and EMILIA~\cite{he2024emilia} are 3.20, 3.33 and 3.22, respectively.

\subsection{Baseline TTS benchmarks}
We present the performance of four different TTS systems, all trained with TITW datasets: (i)~TransformerTTS~\cite{li2019transformertts} with ParallelWaveGAN~\cite{yamamoto2020pwg}; (ii)~GradTTS-DiffWave~\cite{popov2021gradtts,kong2021diffwave}; (iii)~VITS~\cite{kim2021vits}; (iv)~MQTTS~\cite{chen2023vector}.
The choice of these baseline models aims to offer a diverse representation of contemporary TTS technologies.
All models were trained with an open-source recipe for reproducibility with detailed recipes at \cite{jung2024spoofceleb}.

Table~\ref{tab:easyset} displays the results for these four TTS models when trained with the TITW-Easy dataset, evaluated under both protocols.
UTMOS and DNSMOS scores of the synthesized speech being comparable with those in Table~\ref{tab:sqa_bona} show that it matches the quality of training data. 
This indicates that systems were successfully trained.
Yet, they struggle with the inherent challenges of the TITW data.  
The WER significantly increases in most cases. 
These baseline performances are further challenged by the TITW-KSUT protocol results, where all performances degrade compared to the TITW-KSKT protocol.

Table~\ref{tab:hardset} compares the performance of models trained on TITW-Easy and TITW-Hard datasets.
Here, we focus on two systems, GradTTS-DiffWave and VITS, as the other two baseline systems failed to converge when trained with TITW-Hard.
We also present the result replacing GradTTS with a mel-spectrogram extracted from the original speech file (i.e., copy synthesis), which serves as the upper bound for the waveform model, DiffWave.
The results consistently confirm that in all cases models trained on TITW-Hard produce speech of lower quality, highlighting the challenging nature of TITW-Hard.

\section{Conclusion and remarks}
\label{sec:conclusion}
We introduce TITW, a new dataset tailored for training, evaluation and benchmarking TTS systems using real-world, in-the-wild speech data. 
TITW responds to the growing trend in TTS research toward noisy-TTS training by leveraging uncontrolled environments.
Through a fully automated processing pipeline applied to VoxCeleb1—chosen for its diverse, YouTube-sourced speech—we ensure scalability and broad accessibility.
Our results demonstrate that four state-of-the-art TTS systems, when trained on TITW-Easy, produce synthetic speech that closely rivals the quality of the training data.
However, our analysis reveals that only modern deep-learning-based TTS systems can effectively utilize TITW, while older statistical or early neural network-based systems struggle.  
Training is also sensitive to data preparation, due to variability in noise, accents, or recording conditions, which might explain why the noisy-TTS training field has emerged only recently.

Beyond technical advancements, TITW’s design carries significant ethical potential. By using an automatic speaker verification data as a source, it supports research into speech deepfake detection, a crucial task for combating the malicious use of synthetic voices.
Consequently, TITW not only enhances TTS development for underrepresented languages lacking high-quality datasets but also bolsters safeguards against generative speech misuse.
We hope that making TITW publicly available will spark further exploration of noisy-TTS training, driving both innovation and ethical responsibility in synthetic speech technology.

\clearpage

\bibliographystyle{IEEEtran}
\bibliography{refs}

\begin{thebibliography}{10}
\providecommand{\url}[1]{#1}
\csname url@samestyle\endcsname
\providecommand{\newblock}{\relax}
\providecommand{\bibinfo}[2]{#2}
\providecommand{\BIBentrySTDinterwordspacing}{\spaceskip=0pt\relax}
\providecommand{\BIBentryALTinterwordstretchfactor}{4}
\providecommand{\BIBentryALTinterwordspacing}{\spaceskip=\fontdimen2\font plus
\BIBentryALTinterwordstretchfactor\fontdimen3\font minus \fontdimen4\font\relax}
\providecommand{\BIBforeignlanguage}[2]{{%
\expandafter\ifx\csname l@#1\endcsname\relax
\typeout{** WARNING: IEEEtran.bst: No hyphenation pattern has been}%
\typeout{** loaded for the language `#1'. Using the pattern for}%
\typeout{** the default language instead.}%
\else
\language=\csname l@#1\endcsname
\fi
#2}}
\providecommand{\BIBdecl}{\relax}
\BIBdecl

\bibitem{le2024voicebox}
M.~Le, A.~Vyas \emph{et~al.}, ``Voicebox: Text-guided multilingual universal speech generation at scale,'' in \emph{Proc. NeurIPS}, 2024.

\bibitem{kim2024p}
S.~Kim, K.~Shih \emph{et~al.}, ``P-flow: a fast and data-efficient zero-shot {TTS} through speech prompting,'' in \emph{Proc. NeurIPS}, 2024.

\bibitem{nguyen2024fregrad}
T.~D. Nguyen, J.-H. Kim \emph{et~al.}, ``Fregrad: Lightweight and fast frequency-aware diffusion vocoder,'' in \emph{Proc. IEEE ICASSP}, 2024.

\bibitem{zhang2023speechtokenizer}
X.~Zhang, D.~Zhang \emph{et~al.}, ``Speechtokenizer: Unified speech tokenizer for speech large language models,'' in \emph{Proc. ICLR}, 2024.

\bibitem{yang2023uniaudio}
D.~Yang, J.~Tian \emph{et~al.}, ``{UniAudio}: An audio foundation model toward universal audio generation,'' in \emph{Proc. ICML}, 2024.

\bibitem{lee2022pvae}
J.-H. Lee, S.-H. Lee \emph{et~al.}, ``{PVAE-TTS}: Adaptive text-to-speech via progressive style adaptation,'' in \emph{Proc. IEEE ICASSP}, 2022.

\bibitem{kharitonov2023speak}
E.~Kharitonov, D.~Vincent \emph{et~al.}, ``Speak, read and prompt: High-fidelity text-to-speech with minimal supervision,'' \emph{Transactions of the Association for Computational Linguistics}, 2023.

\bibitem{kimclam}
J.~Kim, K.~Lee \emph{et~al.}, ``Clam-{TTS}: Improving neural codec language model for zero-shot text-to-speech,'' in \emph{Proc. ICLR}, 2024.

\bibitem{yamagishi2010thousands}
J.~Yamagishi, B.~Usabaev \emph{et~al.}, ``Thousands of voices for {{HMM-based}} speech synthesis--{{Analysis}} and application of {{TTS}} systems built on various {{ASR}} corpora,'' \emph{IEEE Transactions on Audio, Speech, and Language Processing}, vol.~18, no.~5, pp. 984--1004, 2010.

\bibitem{karhilaNoise2014}
R.~Karhila, U.~Remes, and M.~Kurimo, ``Noise in {{HMM-Based Speech Synthesis Adaptation}}: {{Analysis}}, {{Evaluation Methods}} and {{Experiments}},'' \emph{IEEE Journal of Selected Topics in Signal Processing}, vol.~8, no.~2, pp. 285--295, Apr. 2014.

\bibitem{zhang2021denoispeech}
C.~Zhang, Y.~Ren \emph{et~al.}, ``{DenoiSpeech}: Denoising text to speech with frame-level noise modeling,'' in \emph{Proc. IEEE ICASSP}, 2021, pp. 7063--7067.

\bibitem{chen2023vector}
L.-W. Chen, S.~Watanabe, and A.~Rudnicky, ``A vector quantized approach for text to speech synthesis on real-world spontaneous speech,'' in \emph{Proc. AAAI}, 2023, pp. 12\,644--12\,652.

\bibitem{wang2024asvspoof}
X.~Wang, H.~Delgado \emph{et~al.}, ``{ASV}spoof 5: Crowdsourced speech data, deepfakes, and adversarial attacks at scale,'' in \emph{Proc. Interspeech}, 2024.

\bibitem{nagrani2017voxceleb}
A.~Nagrani, J.~S. Chung, and A.~Zisserman, ``{VoxCeleb}: A large-scale speaker identification dataset,'' in \emph{Proc. Interspeech}, 2017.

\bibitem{mai2023warning}
K.~T. Mai, S.~Bray \emph{et~al.}, ``Warning: humans cannot reliably detect speech deepfakes,'' \emph{PLOS One}, vol.~18, no.~8, pp. 1--20, 2023.

\bibitem{jung2022sasv}
J.-w. Jung, H.~Tak \emph{et~al.}, ``{SASV} 2022: The first spoofing-aware speaker verification challenge,'' in \emph{Proc. Interspeech}, 2022, pp. 2893--2897.

\bibitem{kominek2004cmuarctic}
J.~Kominek and A.~W. Black, ``The {CMU} arctic speech databases,'' in \emph{Proc. Interspeech}, 2004.

\bibitem{yamagishiCSTR2019}
J.~Yamagishi, C.~Veaux, and K.~MacDonald, ``{{CSTR VCTK Corpus}}: {{English Multi-speaker Corpus}} for {{CSTR Voice Cloning Toolkit}} (version 0.92),'' 2019.

\bibitem{hunt1996unit}
A.~J. Hunt and A.~W. Black, ``Unit selection in a concatenative speech synthesis system using a large speech database,'' in \emph{Proc. IEEE ICASSP}, 1996.

\bibitem{tokuda2013speech}
K.~Tokuda, Y.~Nankaku \emph{et~al.}, ``Speech synthesis based on hidden {{Markov}} models,'' \emph{Proceedings of the IEEE}, vol. 101, no.~5, pp. 1234--1252, 2013.

\bibitem{ljspeech17}
K.~Ito and L.~Johnson, ``The {LJ} speech dataset,'' \url{https://keithito.com/LJ-Speech-Dataset/}, 2017.

\bibitem{pratap2020mls}
V.~Pratap, Q.~Xu \emph{et~al.}, ``{MLS}: A large-scale multilingual dataset for speech research,'' in \emph{Proc. Interspeech}, 2020.

\bibitem{zen2019libritts}
H.~Zen, V.~Dang \emph{et~al.}, ``Libri{TTS}: A corpus derived from librispeech for text-to-speech,'' in \emph{Proc. Interspeech}, 2019.

\bibitem{wang2023neural}
C.~Wang, S.~Chen \emph{et~al.}, ``Neural codec language models are zero-shot text to speech synthesizers,'' \emph{arXiv preprint arXiv:2301.02111}, 2023.

\bibitem{shennaturalspeech}
K.~Shen, Z.~Ju \emph{et~al.}, ``{NaturalSpeech} 2: Latent diffusion models are natural and zero-shot speech and singing synthesizers,'' in \emph{Proc. ICLR}, 2024.

\bibitem{he2024emilia}
H.~He, Z.~Shang \emph{et~al.}, ``Emilia: An extensive, multilingual, and diverse speech dataset for large-scale speech generation,'' \emph{arXiv preprint arXiv:2407.05361}, 2024.

\bibitem{jung2024spoofceleb}
J.-w. Jung, Y.~Wu \emph{et~al.}, ``Spoofceleb: Speech deepfake detection and sasv in the wild,'' \emph{IEEE Open Journal of Signal Processing}, 2025.

\bibitem{WhisperX-Bain2023}
M.~Bain, J.~Huh \emph{et~al.}, ``{WhisperX}: Time-accurate speech transcription of long-form audio,'' in \emph{Proc. Interspeech}, 2023, pp. 4489--4493.

\bibitem{Whisper-Radford2023}
A.~Radford, J.~W. Kim \emph{et~al.}, ``Robust speech recognition via large-scale weak supervision,'' in \emph{Proc. ICML}, 2023, pp. 28\,492--28\,518.

\bibitem{OWSMv3-Peng2024}
Y.~Peng, J.~Tian \emph{et~al.}, ``{OWSM} v3.1: Better and faster open {Whisper}-style speech models based on e-branchformer,'' in \emph{Proc. Interspeech}, 2024, pp. 352--356.

\bibitem{Denoiser-Dfossez2020}
A.~Défossez, G.~Synnaeve, and Y.~Adi, ``Real time speech enhancement in the waveform domain,'' in \emph{Proc. Interspeech}, 2020, pp. 3291--3295.

\bibitem{DNSMOS-Reddy2021}
C.~K. Reddy, V.~Gopal, and R.~Cutler, ``{DNSMOS}: A non-intrusive perceptual objective speech quality metric to evaluate noise suppressors,'' in \emph{Proc. IEEE ICASSP}, 2021, pp. 6493--6497.

\bibitem{DNSMOS_P835-Reddy2022}
------, ``{DNSMOS} {P}.835: A non-intrusive perceptual objective speech quality metric to evaluate noise suppressors,'' in \emph{Proc. IEEE ICASSP}, 2022, pp. 886--890.

\bibitem{fairbanks1960voice}
G.~Fairbanks, ``Voice and articulation drillbook,'' 1960.

\bibitem{benoit1996sus}
C.~Beno{\^\i}t, M.~Grice, and V.~Hazan, ``The {SUS} test: A method for the assessment of text-to-speech synthesis intelligibility using semantically unpredictable sentences,'' \emph{Speech Communication}, vol.~18, no.~4, pp. 381--392, 1996.

\bibitem{king2014measuring}
S.~King, ``Measuring a decade of progress in text-to-speech,'' \emph{Loquens}, vol.~1, no.~1, pp. e006--e006, 2014.

\bibitem{kubichek1993mel}
R.~Kubichek, ``Mel-cepstral distance measure for objective speech quality assessment,'' in \emph{Proc. IEEE pacific rim conference on communications computers and signal processing}, 1993.

\bibitem{saeki2022utmos}
T.~Saeki, D.~Xin \emph{et~al.}, ``{UTMOS}: {UTokyo}-{SaruLab} system for {VoiceMOS} challenge 2022,'' in \emph{Proc. Interspeech}, 2022.

\bibitem{radford2023robust}
A.~Radford, J.~W. Kim \emph{et~al.}, ``Robust speech recognition via large-scale weak supervision,'' in \emph{Proc. ICML}.\hskip 1em plus 0.5em minus 0.4em\relax PMLR, 2023.

\bibitem{shi2024codec}
J.~Shi, J.~Tian \emph{et~al.}, ``{ESPnet-Codec}: Comprehensive training and evaluation of neural codecs for audio, music, and speech,'' in \emph{Proc. SLT}, 2024.

\bibitem{veaux201x_cstr_vctk}
C.~Veaux, J.~Yamagishi, and K.~MacDonald, ``{CSTR VCTK Corpus: English Multi-speaker Corpus for CSTR Voice Cloning Toolkit},'' The Centre for Speech Technology Research (CSTR), University of Edinburgh, 2019.

\bibitem{li2019transformertts}
N.~Li, S.~Liu \emph{et~al.}, ``Neural speech synthesis with transformer network,'' in \emph{Proc. AAAI}, 2019, pp. 6706--6713.

\bibitem{yamamoto2020pwg}
R.~Yamamoto, E.~Song, and J.-M. Kim, ``Parallel {WaveGAN}: A fast waveform generation model based on generative adversarial networks with multi-resolution spectrogram,'' in \emph{Proc. IEEE ICASSP}, 2020, pp. 6199--6203.

\bibitem{popov2021gradtts}
V.~Popov, I.~Vovk \emph{et~al.}, ``Grad-{TTS}: A diffusion probabilistic model for text-to-speech,'' in \emph{Proc. ICML}, 2021, pp. 8599--8608.

\bibitem{kong2021diffwave}
Z.~Kong, W.~Ping \emph{et~al.}, ``{DiffWave}: A versatile diffusion model for audio synthesis,'' in \emph{Proc. ICLR}, 2021.

\bibitem{kim2021vits}
J.~Kim, J.~Kong, and J.~Son, ``Conditional variational autoencoder with adversarial learning for end-to-end text-to-speech,'' in \emph{Proc. ICML}, 2021, pp. 5530--5540.

\end{thebibliography}
\end{document}